\documentstyle[eqsecnum,prd,aps,twocolumn]{revtex}
\input{epsf}
\tighten
\begin{document}

\twocolumn[\hsize\textwidth\columnwidth\hsize
\csname@twocolumnfalse\endcsname
\title{Optical geometry for gravitational collapse and
Hawking radiation}

\author{Sebastiano
Sonego,$^{1,a}$ Joachim Almergren,$^{2,b}$ and Marek A.\
Abramowicz$^{2,3,4,c}$}
\address{${^1}$Universit\`a di Udine,
Via delle Scienze 208, 33100 Udine, Italy}
\address{$^2$International School for Advanced
Studies, Via Beirut 2--4, 34014 Trieste, Italy}
\address{$^3$Department of Astronomy and Astrophysics,
Chalmers University of Technology, 41296 G\"oteborg, Sweden}
\address{$^4$International Centre for Theoretical Physics,
Strada Costiera 11, 34014 Trieste, Italy}

\maketitle

\begin{abstract} The notion of
optical geometry, introduced more than twenty years ago as a
formal tool in quantum field theory on a static background,
has recently found several applications to the study of
physical processes around compact objects. In this paper we
define optical geometry for spherically symmetric
gravitational collapse, with the purpose of extending the
current formalism to physically interesting spacetimes which
are not conformally static. The treatment is fully general
but, as an example, we also discuss the special case of the
Oppenheimer-Snyder model. The analysis of the late time
behaviour shows a close correspondence between the structure
of optical spacetime for gravitational collapse and that of
flat spacetime with an accelerating boundary. Thus, optical
geometry provides a natural physical interpretation for
derivations of the Hawking effect based on the ``moving
mirror analogy.'' Finally, we briefly discuss the issue of
back-reaction in black hole evaporation and the information
paradox from the perspective of optical geometry.
\vskip.5cm
\noindent PACS number(s): 04.62.+v, 04.70.Dy, 04.70.Bw,
04.90.+e
\end{abstract}

]

\newcommand{\beq}{\begin{equation}}
\newcommand{\eeq}{\end{equation}}
\newcommand{\lab}{\label}
\newcommand{\dd}{{\rm d}}
\newcommand{\ee}{{\rm e}}
\newcommand{\nab}{\nabla\!}

\narrowtext


\section{Introduction}

\lab{1}

A conformally static spacetime\footnote{We adopt the
notations and conventions of Ref.\ \cite{wald}.} $({\cal
M},g_{ab})$ admits a privileged congruence of timelike
curves, corresponding to the flow lines of conformal Killing
time $t$. Consequently, one can define a family of
privileged observers with four-velocity
$n^a=\eta^a/(-\eta_b\eta^b)^{1/2}$, where $\eta^a$ is the
conformal Killing vector field. The set of these observers
can be thought of as a generalization of the Newtonian
concept of a rest frame. Their acceleration can be expressed
as the projection of a gradient,
\beq n^b\nab_b
n_a={h_a}^b\nab_b\Phi
\lab{A}\eeq
(see Appendix A for a proof), where ${h_a}^b={\delta_a}^b+
n_a n^b$ and
\beq \Phi={1\over 2}\,\ln\left(-\eta_a\eta^a\right)\;;
\lab{Phi}\eeq
thus, $\Phi$ is a suitable general-relativistic counterpart
of the gravitational potential \cite{gsv}. One can define
the ultrastatic \cite{fulling} metric $\tilde{g}_{ab}=
(-\eta_c\eta^c)^{-1}g_{ab}=\ee^{-2\Phi}g_{ab}$, which can be
written as $\tilde{g}_{ab}=-\nab_a t\nab_b
t+\tilde{h}_{ab}$, where
$\tilde{h}_{ab}=\ee^{-2\Phi}h_{ab}$. The hypersurfaces
$t=\mbox{const}$ of ${\cal M}$ are all diffeomorphic to some
three-dimensional manifold $\cal S$. If the spacetime is
static, it follows from Fermat's principle that light rays
coincide with the geodesics on $\cal S$ according to
$\tilde{h}_{ab}$ \cite{og}. For this reason,
$\tilde{g}_{ab}$ is called the {\em optical metric\/}
\cite{og}, and $({\cal S},\tilde{h}_{ab})$ the {\em optical
space\/}. We shall also refer to the family of preferred
observers $n^a$ as the {\em optical frame\/}.

There is a simple operational definition of the optical
metric. Suppose that all the observers $n^a$ agree to
construct a set of synchronized devices that measure the
Killing time $t$. (Of course, these ``clocks'' will not
agree with those based on local physical processes --- e.g.,
on atomic transitions --- but this is totally irrelevant for
the following argument.) Then, they use light signals
according to a radar procedure, and define the distance
between two points $P,Q\in {\cal S}$ as
$t_{\scriptscriptstyle PQP}/2$, where $t_{\scriptscriptstyle
PQP}= t_{\scriptscriptstyle QPQ}$ is the lapse of Killing
time corresponding to the round trip of the signal between
the observers based at $P$ and $Q$.\footnote{There is a
one-to-one correspondence between conformally static
observers and points of $\cal S$.} In this way, they
attribute the metric $\tilde{h}_{ab}$ to $\cal S$.

\begin{figure}
\centerline{\epsfysize=7cm \epsfbox{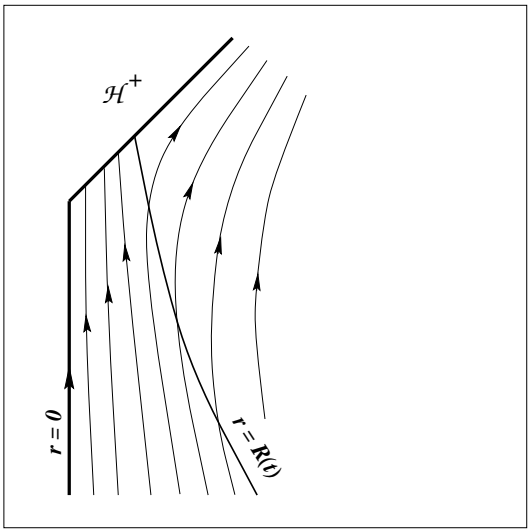}}
\vskip.4cm
\caption{Kruskal diagram for the collapse of a spherically
symmetric shell of matter; only the region outside the
horizon is shown. The natural ``rest frame'' inside the
shell does not match satisfactorily with the Schwarzschild
frame, because the behaviours of the two frames near the
horizon are very different.}
\label{fig1}
\end{figure}

The notion of optical geometry has recently received
considerable attention as a powerful tool in general
relativity \cite{appl,aabgs}. It is thus important to
investigate to which extent it can be generalized to
spacetimes that are not conformally static. One proposal in
this direction \cite{a93} appears mainly formal, and is
probably not sufficient in order to determine $n^a$ and
$\tilde{g}_{ab}$ in a unique way for an arbitrary spacetime
\cite{sm}. It is perhaps more helpful to focus on specific
situations, that may provide one with additional, physically
motivated, hints. In the present article we study the case
of a spacetime that describes the gravitational collapse of
a spherically symmetric configuration of matter. This
problem is interesting for two reasons. First, it represents
one of the simplest cases in which the property of conformal
staticity does not hold. This is particularly evident if we
consider a situation in which the collapsing matter is
concentrated on an infinitely thin shell. In this case, the
spacetime is composed of two regions, corresponding to the
interior and the exterior of the shell, joined through a
timelike hypersurface which represents the history of the
shell. Both these regions are static when considered
separately, their metrics being the Minkowski and the
Schwarzschild ones, respectively. However, the fields $n^a$
associated with these two metrics do not match in a
satisfactory way across the surface of the shell (see Fig.\
\ref{fig1}). In particular, the horizon is a singular locus
for the Schwarzschild frame, but is perfectly regular for
the Minkowski observers. This very different behaviour
prevents one from considering a single frame that reduces to
the Schwarzschild and the Minkowski one, respectively,
outside and inside the shell. The failure can be seen as a
consequence of the fact that the spacetime is not
conformally static in any region containing the
shell.\footnote{See Ref.\ \cite{bondi-rindler} for
considerations related to this point.} Indeed, independently
of its specific properties, the shell represents a
non-stationary boundary between two static regions. A second
motivation for studying this class of spacetimes is that
they lead to the Hawking effect \cite{hawking}. Given the
success of optical geometry in discussing complicated
physical phenomena, one expects that it might give new
insight about the process of black hole evaporation. Indeed,
this appears to be the case, as we shall see.

The structure of the paper is the following. In the next
section we present a general construction of the optical
geometry for an arbitrary matter configuration undergoing
spherical collapse. Section \ref{4} is devoted to the
analysis of some features that become universal (i.e.,
model-independent) at late times. In Sec.\ \ref{3} we
consider a very simple particular case, the
Oppenheimer-Snyder dust model. In Sec.\ \ref{5} we argue
that the optical geometry picture of collapse is the natural
framework for derivations of the Hawking effect based on the
``moving mirror analogy,'' and that it gives useful insight
about the issue of black hole evaporation and the
information paradox. Section \ref{6} contains a summary of
the results, together with some final comments and outlines
for future investigations.


\section{General construction}

\lab{2}

The metric of any spherically symmetric spacetime can be
written as
\beq g=-\alpha(t,r)\dd t^2+\beta(t,r)\dd
r^2+r^2\left(\dd\theta^2+\sin^2\theta\,\dd\varphi^2\right)\;,
\lab{sph}\eeq
with $\alpha$ and $\beta$ positive functions (see, e.g.,
Ref.\ \cite{mtw}, pp.\ 616--617). In the following we
consider situations where matter is confined to a region
$r\leq R(t)$, with $R(t)$ a known function (the ``radius of
the star''). For $r>R(t)$, we assume that the spacetime is
empty. However, the treatment can be easily extended to
include more general types of collapse
--- e.g., of electrically charged configurations \cite{lrs}.
According to Birkhoff's theorem, the metric in the external
region is the Schwarzschild one, thus we have
$\alpha(t,r)=\beta(t,r)^{-1}=C(r):=1-2M/r$ for $r>R(t)$.

\begin{figure}
\centerline{\epsfysize=7cm \epsfbox{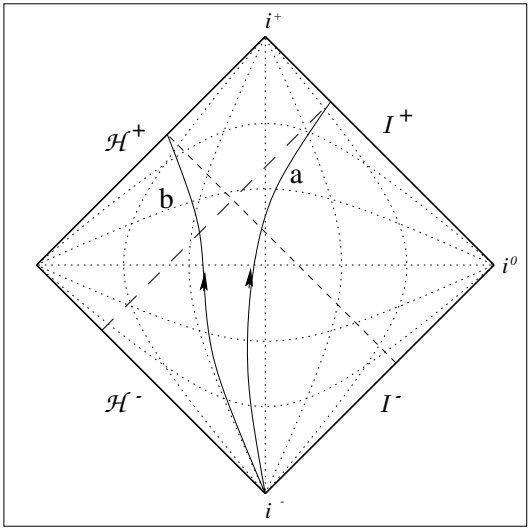}}
\vskip.4cm
\caption{Penrose-Carter diagram of the $r>2M$ region of
Schwarzschild spacetime.  Lines of constant $t$ and of
constant $r$ are drawn, as well as the worldlines of two
observers, one escaping to $r\to +\infty$ (worldline a), the
other entering in the black hole (worldline b).}
\label{fig2}
\end{figure}

\begin{figure}
\centerline{\epsfysize=7cm \epsfbox{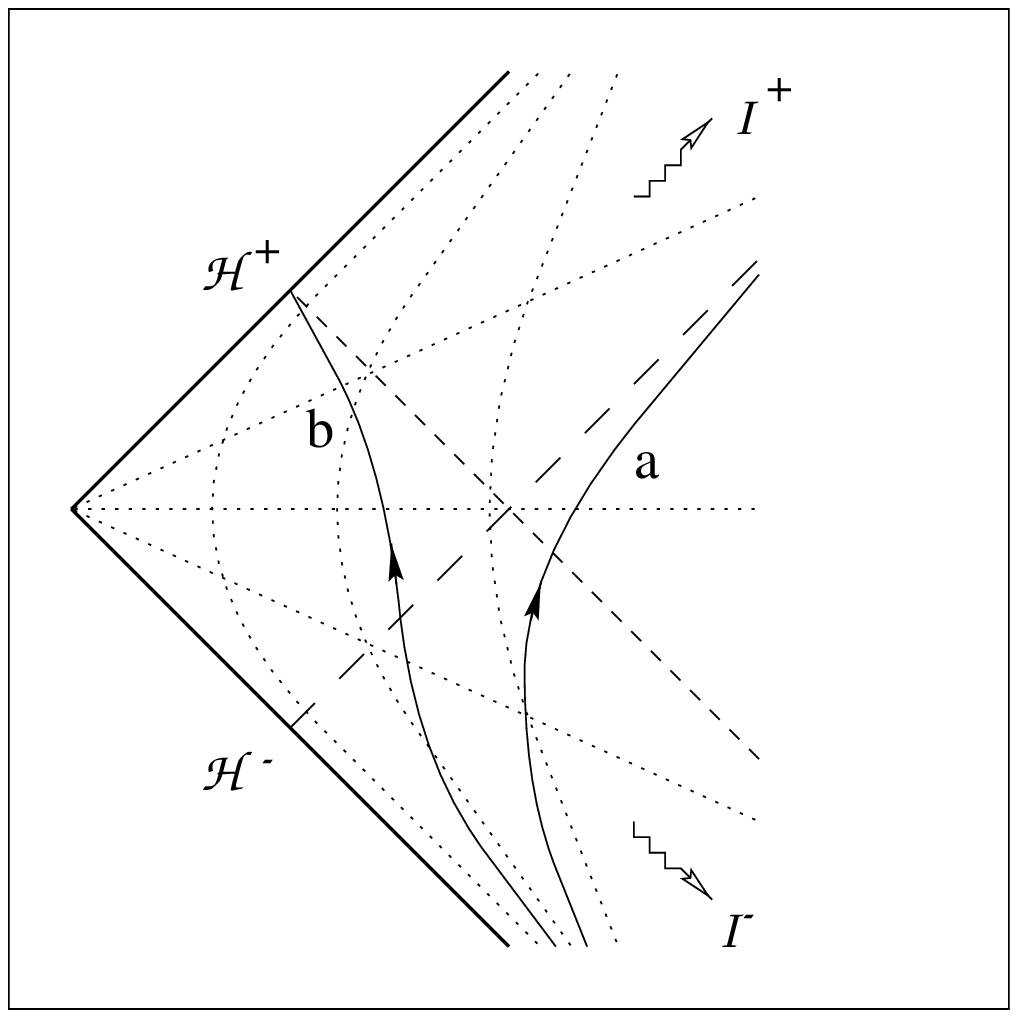}}
\vskip.4cm
\caption{Kruskal diagram of the $r>2M$ region of
Schwarzschild spacetime. This figure can be regarded as a
``partial decompactification'' of Fig.\ \ref{fig2}, wherein
${\cal I}^\pm$ are ``sent to infinity'' and lines
$r=\mbox{const}$ are ``straightened up'' for large values of
$r$.}
\label{fig3}
\end{figure}

\begin{figure}
\centerline{\epsfysize=7cm \epsfbox{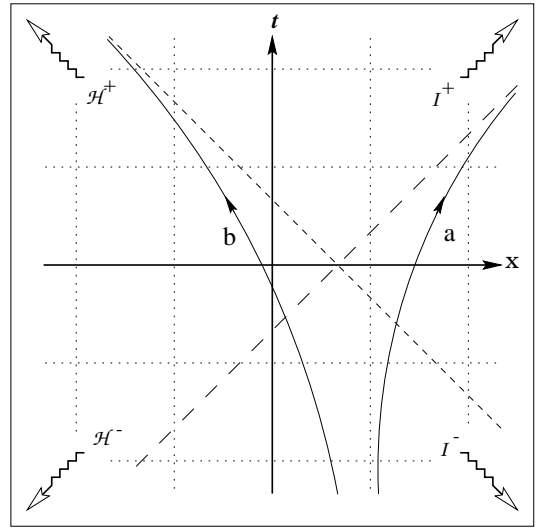}}
\vskip.4cm
\caption{The $r>2M$ region of Schwarzschild
spacetime in $(t,x)$ coordinates. This is how a
$\theta=\mbox{const}$, $\varphi=\mbox{const}$ section of
Schwarzschild spacetime appears when the optical metric
$\tilde{g}_{ab}$ is used. The worldlines $r=\mbox{const}$ of
the privileged observers have been completely straightened,
and the horizons ${\cal H}^\pm$ have been ``sent to
infinity.'' Notice that the observers a and b appear
accelerated in this diagram.}
\label{fig4}
\end{figure}

In this case, the ``rest frame'' $n^a$ outside the star is
just made of the Schwarzschild static observers,
$n^\mu=C(r)^{-1/2}\delta_t^\mu$, and the optical geometry is
$\tilde{g}_{ab}=C(r)^{-1}g_{ab}$. Introducing the
Regge-Wheeler ``tortoise'' coordinate $x$, such that $\dd
x:=C(r)^{-1}\dd r$, we have
\beq \tilde{g}=-\dd t^2+\dd x^2+\tilde{r}(x)^2
\left(\dd\theta^2+\sin^2\theta\,\dd\varphi^2\right)\;,
\lab{optschw}\eeq
where $\tilde{r}:=C(r)^{-1/2}r$. The Regge-Wheeler
coordinate has therefore a very simple geometrical meaning
in the optical space: It expresses directly the value of
radial distances on $({\cal S},\tilde{h}_{ab})$. Notice
that, as far as purely radial motions are concerned, the
optical metric (\ref{optschw}) gives the same line element
as Minkowski spacetime. In particular, no event horizon is
present, because the conformal transformation from $g_{ab}$
to $\tilde{g}_{ab}$ ``sends'' the Schwarzschild horizon
$r=2M$ to infinity. In fact, for a spacetime with metric
$\tilde{g}_{ab}$ the points with $r=2M$ belong to the null
infinity, and the conformal rescaling that carries $g_{ab}$
into $\tilde{g}_{ab}$ can be compared to the
``decompactification'' of a Penrose-Carter diagram, as it is
evident from Figs.\ \ref{fig2}--\ref{fig4}.

To define ``natural'' observers inside the star is not so
easy. In general, the metric in the internal region is not
conformally static, and one cannot thus apply the
construction based on the timelike conformal Killing vector
field, outlined at the beginning of Sec.\ \ref{1}. However,
even when such a field exists it does not necessarily
produce a satisfactory family of internal observers. This
point can be clarified by considering again the example of a
collapsing shell of matter. Inside the shell the spacetime
is flat, by Birkhoff's theorem; therefore, it would seem
obvious to choose inertial observers at fixed distances with
respect to the centre of the shell, in order to define a
``rest frame.'' But such observers are not the natural
continuation inside the shell of the Schwarzschild static
ones, defined outside. This can be seen by noticing that the
horizon ${\cal H}^+$ is infinitely ahead in the future for
the Schwarzschild observers, but not for those at rest with
respect to the centre of the shell. Similarly, the
Schwarzschild observers ``crowd'' near ${\cal H}^+$, unlike
the internal ones (see Fig.\ \ref{fig1}). Thus, using the
Schwarzschild and the inertial frames would lead to
ill-behaved optical metric $\tilde{h}_{ab}$ and potential
$\Phi$.

\begin{figure}
\centerline{\epsfysize=7cm \epsfbox{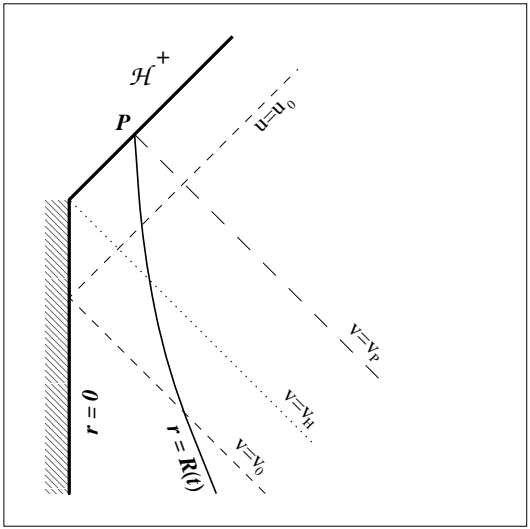}}
\vskip.4cm
\caption{Light rays in a Kruskal diagram for
spherically symmetric collapse. The outgoing ray $u=u_0$ is
generated by the incoming one, $v=v_0$, after it has crossed
the centre of the star. It appears reflected because the
diagram is in polar coordinates.}
\label{fig5}
\end{figure}

\begin{figure}
\centerline{\epsfysize=7cm \epsfbox{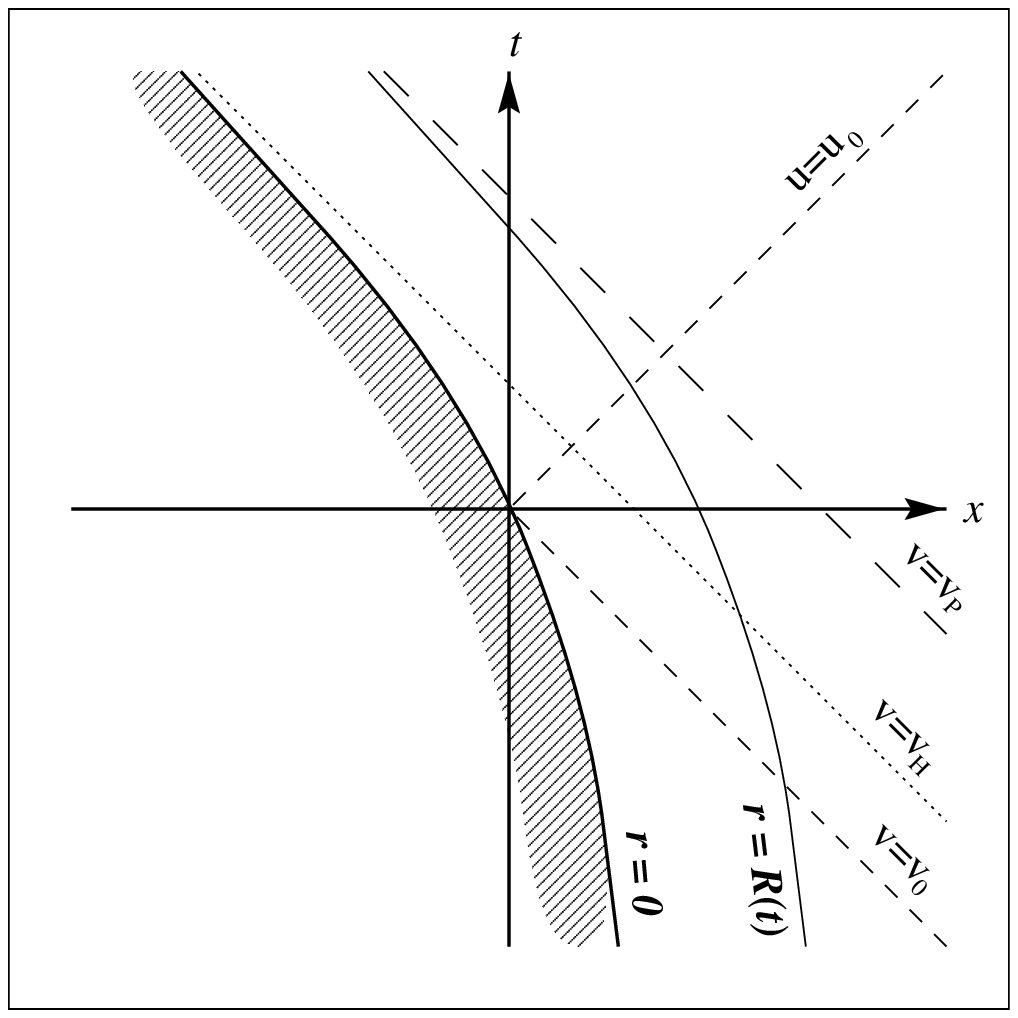}}
\vskip.4cm
\caption{Collapse in $(t,x)$ coordinates.  Both the
surface and the centre of the star appear accelerated, and
their worldlines are asymptotically parallel to each other.
The radius of the star in the optical geometry approaches
the value $v_{\scriptscriptstyle P}-v_{\scriptscriptstyle
H}$ as $t\to +\infty$.  Only the part of the $(t,x)$ plane
above the curve $r=0$ is physically meaningful.}
\label{fig6}
\end{figure}

\begin{figure}
\centerline{\epsfysize=7cm \epsfbox{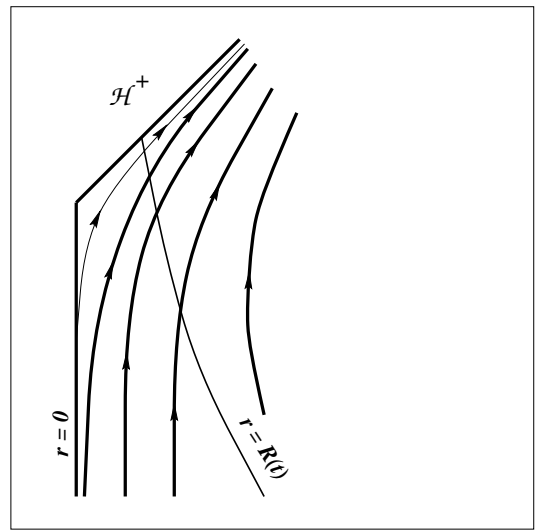}}
\vskip.4cm
\caption{Regular extension of the Schwarzschild frame.}
\label{fig7}
\end{figure}

Before constructing explicitly an extension of the
Schwarzschild frame that does not suffer from these
problems, let us present a graphical discussion of some of
its properties. Basically, we are looking for a continuation
of the coordinates $t$ and $x$ inside matter, such that $\dd
t/\dd x=\pm 1$ for light signals and fundamental observers
are located at $x=\mbox{const}$. In a $(t,x)$ diagram, the
surface of the star is represented by a line like b in Fig.\
\ref{fig4}, so that we have still only to establish how the
centre $r=0$ looks like. To this end, it is convenient to
consider the Kruskal diagram in Fig.\ \ref{fig5}, which
shows three incoming radial light rays (or spherical
wavefronts). The ray at $v=v_0$ simply passes through the
centre of the star and is then converted into an outgoing
signal, $u=u_0$. The ray $v=v_{\scriptscriptstyle H}$
reaches $r=0$ just on the horizon and then turns into a null
generator of ${\cal H}^+$. For $v>v_{\scriptscriptstyle H}$,
all incoming signals enter the black hole region; in
particular, the ray $v=v_{\scriptscriptstyle P}$ does so
exactly when the surface of the star crosses the horizon
(event $P$ in Fig.\ \ref{fig5}). Since light signals are
still represented by straight lines at $\pm 45^\circ$ in the
$(t,x)$ diagram, and $v_0$, $v_{\scriptscriptstyle H}$,
$v_{\scriptscriptstyle P}$ all have finite values, it
follows that the centre $r=0$ must correspond to a line that
becomes asymptotically parallel to the one representing the
surface of the star, as shown in Fig.\ \ref{fig6}. Since
fundamental observers are represented by vertical lines
$x=\mbox{const}$ in the $(t,x)$ plane, it is easy to see
that their qualitative behaviour in a Kruskal diagram is the
one shown in Fig.\ \ref{fig7}. Their worldlines now
accumulate along the whole ${\cal H}^+$ and match regularly
across the surface of the star. These conditions guarantee
that when the metric $\tilde{g}_{ab}$ is used, ${\cal H}^+$
becomes a regular portion of the future asymptotic null
infinity.

Let us now proceed to construct $n^a$ analytically.  It is
convenient to introduce new coordinates $(\tau,\xi)$ in the
internal region $r<R(t)$, such that
\beq -\alpha(t,r)\dd t^2+\beta(t,r)\dd r^2=
\gamma(\tau,\xi)\left(-\dd\tau^2+\dd\xi^2\right)\;,
\lab{2D}\eeq
with $\gamma$ a positive function. (Such coordinates always
exist, because all two-dimensional spacetimes are
conformally flat.) Of course, $t$ and $r$ become now
functions $t(\tau,\xi)$ and $r(\tau,\xi)$ of the new
coordinates. Thus, the internal metric reads
\beq g=\gamma(\tau,\xi)\left(-\dd\tau^2+\dd\xi^2\right)+
r(\tau,\xi)^2
\left(\dd\theta^2+\sin^2\theta\,\dd\varphi^2\right)\;.
\lab{int}\eeq
Since both $t$ and $\tau$ are timelike coordinates, the
history of a point with $\theta=\mbox{const}$,
$\varphi=\mbox{const}$ on the surface of the star consists
of a sequence of events ordered in $\tau$. Therefore, in
terms of the internal coordinates, the equation of the
surface can be written as $\xi=\Xi(\tau)$, obtained by
solving the equation $r(\tau,\xi)=R(t(\tau,\xi))$ with
respect to $\xi$.

The spacetime metric must be continuous across the surface
of the star. Thus, the external metric (\ref{sph}) and the
internal one, given by (\ref{int}), must agree in the
evaluation of the spacetime interval between two events that
occur on the star's surface. Let us consider two such
events, labeled, in the internal coordinates, by
$(t,R(t),\theta,\varphi)$ and $(t+\delta t,R(t+\delta
t),\theta,\varphi)\approx (t+\delta t,R(t)+R'(t)\delta
t,\theta,\varphi)$, where a prime denotes the derivative of
a function with respect to its argument. Similarly, in
external coordinates we have, for the same events,
$(\tau,\Xi(\tau),\theta,\varphi)$ and
$(\tau+\delta\tau,\Xi(\tau+\delta\tau),\theta,\varphi)\approx
(\tau+\delta\tau,\Xi(\tau)+
\Xi'(\tau)\delta\tau,\theta,\varphi)$. Replacing in Eqs.\
(\ref{sph}) and (\ref{int}), and equating the outcomes by
continuity, we obtain a differential relation between $t$
and $\tau$ at the surface of the star,
\begin{eqnarray}
\gamma(\tau,\Xi(\tau))&&\left[1-\Xi'(\tau)^2\right]
\dd\tau^2\nonumber\\
&&=\left[C(R(t))-R'(t)^2/C(R(t))\right]\dd t^2\;.
\lab{dtdtau}
\end{eqnarray}
Integrating Eq.\ (\ref{dtdtau}) gives a relationship
$\tau=f(t)$ between the values of $t$ and $\tau$ at the
surface.

The form (\ref{int}) of the internal metric is convenient
because it allows one to readily define null coordinates
$(U,V)$,
\beq U=\tau-\xi+U_0\;,
\lab{U}\eeq
\beq V=\tau+\xi+V_0\;,
\lab{V}\eeq
where $U_0$ and $V_0$ are arbitrary constants. The
coordinates $U$ and $V$ have the usual physical meaning: The
locus $U=\mbox{const}$ in spacetime is the history of an
outgoing spherical wavefront of light, while
$V=\mbox{const}$ represents an incoming one. If we introduce
null coordinates $(u,v)$ in the outside region as
\beq u=t-x\;,
\lab{u}\eeq
\beq v=t+x\;,
\lab{v}\eeq
we have that an outgoing spherical wavefront is described,
inside the star, by the equation $U=\mbox{const}$ and,
outside the star, by $u=\mbox{const}$. Therefore, one can
establish a one-to-one correspondence $U(u)$ between the
values of $U$ and $u$, defining $U(u)$ as the internal
$U$-label of the wavefront which, outside, is labeled by
$u$. Similarly, one can define a function $V(v)$. The
explicit form of $U(u)$ can be obtained by solving with
respect to $t$ the equation
\beq t-x(R(t))=u
\lab{eqfort}\eeq
and then substituting the result into
\beq U=f(t)-\Xi(f(t))+U_0\;.
\lab{U(u)}\eeq
Analogously, $V(v)$ is obtained by replacing the solution of
\beq t+x(R(t))=v
\lab{eqforv}\eeq
into
\beq V=f(t)+\Xi(f(t))+V_0\;.
\lab{V(v)}\eeq

The functions $U(u)$ and $V(v)$ can be used to extend the
coordinates $(u,v)$, hence $(t,x)$, also inside the star. It
is sufficient to invert them, getting $u$ and $v$ as
functions of $U$ and $V$, respectively, and then {\em
define\/} $t$ and $x$ as
\beq t:={1\over 2}\,\left(v(V)+u(U)\right)\;,
\lab{t}\eeq
\beq x:={1\over 2}\,\left(v(V)-u(U)\right)\;.
\lab{x}\eeq

In terms of $t$ and $x$, the internal metric (\ref{int})
takes the form
\beq g=\gamma U'V'\left(-\dd t^2+\dd x^2\right)+r^2
\left(\dd\theta^2+\sin^2\theta\,\dd\varphi^2\right)\;,
\lab{int'}\eeq
where $U':=\dd U/\dd u$, $V':=\dd V/\dd v$, and all the
functions are implicitly supposed to be expressed in terms
of $t$ and $x$. Now we choose
\beq \Phi={1\over 2}\,\ln\left(\gamma U'V'\right)\;,
\lab{Phiint}\eeq
so that the internal optical metric reads
\beq \tilde{g}=-\dd t^2+\dd x^2+{r^2\over\gamma U'V'}\,
\left(\dd\theta^2+\sin^2\theta\,\dd\varphi^2\right)\;.
\lab{optint}\eeq
In $(t,r)$ coordinates, the optical frame $n^a$ has
components $n^\mu=\left(\gamma
U'V'\right)^{-1/2}\delta^\mu_t$, i.e., in $(\tau,\xi)$
coordinates,
\beq n^\mu={1\over 2(\gamma U'V')^{1/2}}\left[
\left(V'+U'\right)\delta^\mu_\tau+\left(V'-U'\right)
\delta^\mu_\xi\right]\;.
\lab{n-int}\eeq
This vector field is a satisfactory extension inside the
star of the static Schwarzschild frame, and it is not
difficult to check that, although the metric (\ref{int'}) is
not conformally static, Eq.\ (\ref{A}) is still satisfied.
Thus, we can continue to interpret $\Phi$ as the
gravitational potential.

The factor $U'V'$ can be computed as follows. Let us
consider two outgoing light rays, corresponding to the
values $U$ and $U+\delta U$, $u$ and $u+\delta u$ of the
coordinates, with $\delta U$ and $\delta u$ very small.
Equations (\ref{eqfort}) and (\ref{U(u)}) give us the
coordinate times $t$ and $t+\delta t$ at which these rays
cross the surface of the star, expressed as functions of
$u$, $u+\delta u$, $U$, and $U+\delta U$. This allows us to
find the coefficients that link $\delta u$ and $\delta U$ to
$\delta t$; eliminating $\delta t$ and taking the limit, we
get
\beq
{\dd U\over\dd u}={1-\Xi'(f(t(u)))\over
1-R'(t(u))/C(R(t(u)))}\,f'(t(u))\;,
\lab{nasty}\eeq
where the function $t(u)$ is implicitly defined by Eq.\
(\ref{eqfort}).  Similarly, one obtains from Eqs.\
(\ref{eqforv}) and (\ref{V(v)}), considering two incoming
light rays,
\beq
{\dd V\over\dd v}={1+\Xi'(f(t(v)))\over
1+R'(t(v))/C(R(t(v)))}\,f'(t(v))\;,
\eeq
where now $t(v)$ is given by Eq.\ (\ref{eqforv}).  Using
Eq.\ (\ref{dtdtau}) we get at the end:
\begin{eqnarray}
U'(u)V'(v)&&=\left[{1-\Xi'(f(t(u)))\over
1+\Xi'(f(t(u)))}\,{1+\Xi'(f(t(v)))\over
1-\Xi'(f(t(v)))}\right.\nonumber\\ &&
\times{1+R'(t(u))/C(R(t(u)))\over
1-R'(t(u))/C(R(t(u)))}\nonumber\\ &&
\times{1-R'(t(v))/C(R(t(v)))\over
1+R'(t(v))/C(R(t(v)))}\nonumber\\ &&
\times{C(R(t(u)))\over
\gamma(f(t(u)),\Xi(f(t(u))))}\nonumber\\
&& \left.\times{C(R(t(v)))\over
\gamma(f(t(v)),\Xi(f(t(v))))}\right]^{1/2}\;.
\lab{horrible}\end{eqnarray}


\section{Asymptotic behaviour}

\lab{4}

Equations (\ref{Phiint})--(\ref{horrible}) give a complete
characterization of the optical geometry inside a collapsing
spherically symmetric star. Of course, since they require an
explicit knowledge of the functions $R(t)$, $\Xi(\tau)$,
$f(t)$, and $\gamma(\tau,\xi)$, one can use them only within
a specific model of collapse --- and even in that case their
integration will usually require numerical methods. It is
therefore remarkable that, at late times (i.e., for $t, u\to
+\infty$), one could establish analytically some features
that are universal, in the sense that they do not depend on
the details of the model.

In the limit $u\to +\infty$, it is easy to circumvent the
nasty expression on the right hand side of Eq.\
(\ref{nasty}) by noticing that, although the relationship
between the functions $U$ and $u$ is singular on ${\cal
H}^+$, $U$ is regularly connected to the Kruskal retarded
null coordinate \cite{wald}
\beq {\cal U}=-\exp(-u/4M)
\lab{kruskalu}\eeq
for all values of $\cal U$; this follows from the fact that
both $(U,V)$ and the Kruskal coordinates are regular at the
surface of the star. Then, two outgoing light rays, labeled
by $U$ and $U+\delta U$ inside the star, are labeled by
$\cal U$ and ${\cal U}+\delta{\cal U}$ outside, with $\delta
U=a({\cal U})\delta{\cal U}$, where $a$ is a regular
positive function that depends on the details of collapse
(i.e., $R$, $R'$, $\Xi$, and $\Xi'$) at the moment when the
light rays cross the surface of the star. For rays near the
horizon, i.e., in the limit $u\to +\infty$, one has ${\cal
U}\approx 0$ and thus $\delta U\approx a(0)\delta{\cal U}$,
which gives the desired asymptotic relation \cite{bd}
\beq {\dd U\over\dd u}\sim {a(0)\over 4M}\exp(-u/4M)\;.
\lab{asympt}\eeq
(Hereafter, we use the notation $f_1\sim f_2$ to express the
fact that two functions have the same asymptotic expression
in some limit, i.e., $\lim f_1/f_2=1$.) The constant
positive factor $a(0)$ is all that remains of the details of
collapse in the limit $u\to +\infty$.

The situation is rather different as far as $\dd V/\dd v$ is
concerned. Of course, considering two incoming light rays
labeled by $V$ and $V+\delta V$ inside the star, and by
$\cal V$ and ${\cal V}+\delta{\cal V}$ outside, where
\beq {\cal V}=\exp(v/4M)
\lab{kruskalv}\eeq
is the Kruskal advanced null coordinate, one can still claim
that $\delta V=b({\cal V})\delta{\cal V}$, where $b$ is a
regular positive function depending on the dynamics of the
star's surface when it is crossed by the light rays.
However, since the interior of the star at ${\cal H}^+$
corresponds to an entire range of values for $V$ and $v$
(the interval $[v_{\scriptscriptstyle
H},v_{\scriptscriptstyle P}]$ in Fig.\ \ref{fig5}), the
function $\dd V/\dd v$, although regular everywhere, has not
a universal dependence on $v$.

The asymptotic form (\ref{asympt}) of $U'(u)$ and the
regular dependence of $V'$ on $v$ are nevertheless
sufficient in order to establish the main properties of
optical geometry during the late stages of collapse. Since
$\gamma$, $U'$, and $V'$ are regular positive functions, the
product $\gamma U'V'$ can simply be written as
\beq \gamma U'V'\sim F(v)\exp(-u/4M)\;,
\eeq
where $F(v)$ is a non-vanishing positive function which
depends on the details of collapse. Of course, outside the
star, $U\equiv {\cal U}$, $V\equiv {\cal V}$, and
$\gamma=(32 M^3/r)\exp(-r/2M)$, so $\gamma U'V'=1-2M/r$ and
$F(v)\sim \exp(-1+v/4M)$.

~From Eq.\ (\ref{n-int}) it is evident that, since $U'\to 0$
when $u\to +\infty$, the optical frame behaves like the one
in Fig.\ \ref{fig7}, because the components $n^\tau$ and
$n^\xi$ tend to become equal near ${\cal H}^+$. The optical
metric (\ref{optint}) and the potential $\Phi$ have the
following asymptotic forms:
\beq \tilde{g}\sim -\dd t^2+\dd x^2+
{4M^2\,\ee^{(t-x)/4M}\over F(t+x)}\left(\dd\theta^2+
\sin^2\theta\,\dd\varphi^2\right)\;;
\eeq
\beq \Phi\sim {x-t\over 8M}+{1\over 2}\,\ln F(t+x)\;.
\eeq
Using Eq.\ (\ref{A}), one can easily compute the
acceleration of the fundamental observers. The only
nonvanishing component is
\begin{equation}
n^a\nab_a n_x={\partial\Phi\over\partial x}={1\over
8M}+{F'(t+x)\over 2F(t+x)}\;,
\end{equation}
which gives, up to the sign, the value of the
``gravitational field'' at late times. It is interesting to
notice that, outside the star, $n^a\nab_a n_x=1/4M$, which
coincides with the surface gravity of the black hole (see
Appendix B for a general proof). On the other hand, inside
the star, the term $F'/2F$ gives a correction to the surface
gravity that varies from place to place on the horizon and
depends on the model of collapse.

Let us now consider a timelike hypersurface with equation
$\xi=F(\tau)$, such that near ${\cal H}^+$ one can write
$\xi\sim\nu\tau+\mbox{const}$, where $\nu$ is a constant
(from the expression (\ref{int}) of the metric, it follows
that $-1<\nu<1$). This is the case, for example, of the
centre of the star, $\xi\equiv 0$, or of the star's surface,
for which we have $\nu=\dd\Xi/\dd\tau$ evaluated at the
horizon. Equations (\ref{U}) and (\ref{V}) give then
$(1+\nu)\dd U\approx (1-\nu)\dd V$. Near ${\cal H}^+$, the
coordinate $v$ is approximately constant on the submanifold
identified by $\xi=F(\tau)$ (for example, in the cases of
the centre and of the surface of the star, it is equal to
$v_{\scriptscriptstyle H}$ and $v_{\scriptscriptstyle P}$,
in the notations of Fig.\ \ref{fig5}), and this relation can
be rewritten as
\beq \dd v\sim A\exp(-u/4M)\dd u\;,
\eeq
where $A$ is a cumulative positive constant and we have used
Eq.\ (\ref{asympt}). Integrating and using Eq.\ (\ref{v}),
we get
\beq t+x\sim \bar{v}-K\exp(-u/4M)\;,
\lab{main}\eeq
where $K:=4MA>0$ and the integration constant $\bar{v}$ is
the advanced time at which the surface $\xi=F(\tau)$ crosses
${\cal H}^+$. Equation (\ref{main}) can be rewritten using
Eq.\ (\ref{u}) as
\beq x\sim -t+\bar{v}-K\exp(-t/2M)
\lab{Main}\eeq
(see also Ref.\ \cite{mtw}, p.\ 869, and Ref.\
\cite{thorne}). This equation expresses the asymptotic
behaviour, in $(t,x)$ coordinates, of any worldline that
crosses ${\cal H}^+$. Of course, it is valid both inside and
outside the star.


\section{Example: The Oppenheimer-Snyder model}

\lab{3}

In order to apply the general techniques developed so far to
a specific case, let us consider the simplest model of a
collapsing star, in which matter is a ball of dust with
uniform density \cite{os}. In this case, the internal
solution is part of a spatially closed Friedman spacetime
(see Ref.\ \cite{mtw}, pp.\ 851--856), so we have
$\gamma(\tau,\xi)=a(\tau)^2/a_0^2$ and
\beq r(\tau,\xi)=a(\tau)\sin(\xi/a_0)\;,
\lab{zero}\eeq
where $0\leq\tau < \pi a_0$ ($\tau=0$ corresponds to the
beginning of collapse), $a_0$ is a constant,
$0\leq\xi\leq\Xi_0$, with $\Xi_0<\pi a_0/2$ corresponding to
the surface of the star, and
\beq a(\tau)=a_0\cos^2(\tau/2a_0)\;.
\lab{o}\eeq
It is convenient to introduce the dimensionless variable
$\sigma:=\tau/2 a_0$. Then, the function $R(t)$ is defined
implicitly by the two equations
\beq R=R_i\cos^2\sigma
\lab{R}\eeq
and
\begin{eqnarray}
t&=&2M\ln\left[{\left({\displaystyle {R_i\over
2M}}-1\right)^{1/2}+
\tan\sigma\over \left({\displaystyle {R_i\over 2M}}
-1\right)^{1/2}-\tan\sigma}\right]\nonumber\\
&+&4M\left({R_i\over
2M}-1\right)^{1/2}\left[\sigma+{R_i\over
4M}\left(\sigma+\sin\sigma\cos\sigma\right)\right]\;,
\nonumber\\
\lab{T}\end{eqnarray}
where $R_i=a_0\sin(\Xi_0/a_0)$ is the radius of the star at
the beginning of collapse. Note that for
$\sigma=\sigma^\ast:=\arccos\sqrt{2M/R_i}$, which
corresponds to the event horizon, one has $t=+\infty$, as it
should be.

In this model one has $\Xi(\tau)=\Xi_0=\mbox{const}$. The
functions $U(u)$ and $V(v)$ can, in principle, be obtained
by the systems
\beq \left. \matrix{u &=& t-X \hfill\cr
U &=& \tau-\Xi_0+U_0
\cr}\right\}
\lab{uU}\eeq
and
\beq \left. \matrix{v &=& t+X \hfill\cr
V &=& \tau+\Xi_0+V_0
\cr}\right\}\;,
\lab{vV}\eeq
respectively, where $X=R+2M\ln\left(R/2M-1\right)$ and both
$R$ and $t$ are expressed in terms of $\tau$, according to
Eqs.\ (\ref{R}) and (\ref{T}). From now on, we shall exploit
the freedom in the constants $U_0$ and $V_0$, choosing
$U_0=-V_0=\Xi_0$, so that $U=V=\tau$ at the surface of the
star. In practice, Eq.\ (\ref{T}) is a transcendental one
for $\tau$ and cannot be inverted. However, we can easily
find the inverse functions $u(U)$ and $v(V)$. Consider an
outgoing spherical wavefront of light, characterized by
$U=\bar{U}=\mbox{const}$ and $u=\bar{u}=
\mbox{const}$, respectively inside and outside the star.
In particular, we must have
$\bar{u}=t(\bar{\tau})-X(\bar{\tau})$, where $\bar{\tau}$ is
the value of $\tau$ at the moment when the wavefront crosses
the surface of the star. But from Eq.\ (\ref{uU}) we have
also $\bar{U}= \bar{\tau}$, so Eqs.\ (\ref{R}) and (\ref{T})
allow us to conclude that, inside the star, the relationship
between $U$ and $u$ is
\begin{eqnarray} u(U)&=&t(\tau=U)-X(\tau=U)\nonumber\\
&=&-4M\ln\left({R_i\over 2M}\cos^2\sigma_{\scriptscriptstyle
U}-1\right)\nonumber\\ &+& 4M\ln\left[\left({R_i\over
2M}-1\right)^{1/2}+
\tan\sigma_{\scriptscriptstyle U}\right]-
R_i\cos^2\sigma_{\scriptscriptstyle U}\nonumber\\
&+&2M\ln\cos^2\sigma_{\scriptscriptstyle
U}+4M\left({R_i\over
2M}-1\right)^{1/2}\sigma_{\scriptscriptstyle U}\nonumber\\
&+&R_i\left({R_i\over
2M}-1\right)^{1/2}\left(\sigma_{\scriptscriptstyle U}+
\sin\sigma_{\scriptscriptstyle U}
\cos\sigma_{\scriptscriptstyle U}\right)\;,
\lab{u(U)}\end{eqnarray}
with $\sigma_{\scriptscriptstyle U}=U/2a_0$. Analogously, we
find
\begin{eqnarray} v(V)&=&t(\tau=V)+X(\tau=V)\nonumber\\
&=& 4M\ln\left[\left({R_i\over 2M}-1\right)^{1/2}+
\tan\sigma_{\scriptscriptstyle V}\right]+
R_i\cos^2\sigma_{\scriptscriptstyle V}\nonumber\\
&+&2M\ln\cos^2\sigma_{\scriptscriptstyle V}+
4M\left({R_i\over 2M}-1\right)^{1/2}
\sigma_{\scriptscriptstyle V}\nonumber\\
&+&R_i\left({R_i\over
2M}-1\right)^{1/2}\left(\sigma_{\scriptscriptstyle V}+
\sin\sigma_{\scriptscriptstyle V}
\cos\sigma_{\scriptscriptstyle V}\right)\;,
\lab{v(V)}\end{eqnarray}
where $\sigma_{\scriptscriptstyle V}=V/2a_0$. Substituting
Eqs.\ (\ref{u(U)}) and (\ref{v(V)}) into Eqs.\ (\ref{t}) and
(\ref{x}), we can express $t$ and $x$ inside the star as
functions of $U$ and $V$.

Along the worldlines of the observers belonging to the
optical frame one has $x=\mbox{const}$, i.e., $\dd u/\dd
v=1$.  Writing Eqs.\ (\ref{u(U)}) and (\ref{v(V)}) in
differential form,
\beq \dd u={R_i^2\cos^4\sigma_{\scriptscriptstyle U}\over
2a_0M}\cdot {{\displaystyle \left({R_i\over
2M}-1\right)^{1/2}}+
\tan\sigma_{\scriptscriptstyle U}\over {\displaystyle
{R_i\over 2M}}\cos^2\sigma_{\scriptscriptstyle U}-1}\,\dd
U\;,
\lab{du/dU}\eeq
\beq \dd v={R_i^2\cos^4\sigma_{\scriptscriptstyle V}\over
2a_0M}\cdot{{\displaystyle \left({R_i\over
2M}-1\right)^{1/2}}-
\tan\sigma_{\scriptscriptstyle V}\over
{\displaystyle {R_i\over 2M}\cos^2\sigma_{\scriptscriptstyle
V}-1}}\,\dd V\;,
\lab{dv/dV}\eeq
we find the slope of these worldlines in $(U,V)$ coordinates:
\begin{eqnarray}
{\dd U\over\dd V}&&= {\left(\cos^2\sigma_{\scriptscriptstyle
U}-{\displaystyle {2M\over
R_i}}\right)\cos^4\sigma_{\scriptscriptstyle V}\over
\left(\cos^2\sigma_{\scriptscriptstyle V}-{\displaystyle
{2M\over R_i}}\right)
\cos^4\sigma_{\scriptscriptstyle U}}\nonumber\\
&&\times{\left({\displaystyle {R_i\over 2M}}-
1\right)^{1/2}-\tan\sigma_{\scriptscriptstyle V}\over
\left({\displaystyle {R_i\over 2M}}-1\right)^{1/2}+
\tan\sigma_{\scriptscriptstyle U}}\;.
\lab{dU/dV}\end{eqnarray}
\narrowtext\noindent
This equation can be integrated numerically for suitable
values of $M$ and $R_i$, and produces diagrams that agree
with our qualitative sketch of Fig.\ \ref{fig7}.

The function $u(U)$ assumes a particularly simple asymptotic
form near the event horizon, in agreement with the general
analysis in Sec.\ \ref{4}. The surface of the star crosses
the horizon when $R=2M$, i.e., at a time $\tau=\tau^\ast$
such that $\cos^2\sigma^\ast=2M/R_i$. Since the horizon is a
null hypersurface with $U=U^\ast=\mbox{const}$, we have also
that $\cos^2\sigma^\ast_{\scriptscriptstyle U}=2M/R_i$.
Then, for $U\to U^\ast$ the first term on the right hand
side of Eq.\ (\ref{u(U)}) dominates and one can write,
asymptotically,
\beq u(U)\sim -4M\ln\left(\sigma^\ast_{\scriptscriptstyle U}
-\sigma_{\scriptscriptstyle U}\right)\;.
\lab{u-asympt}\eeq
Consistently, one finds from Eq.\ (\ref{du/dU}) that $\dd
u/\dd U\sim -4M/(U-U^\ast)$; in fact, this relationship is
not restricted to the present model and holds for a generic
collapse, as it follows from Eq.\ (\ref{asympt}). Since
$v(V)$ remains finite, we have also
\beq t(U,V)\sim -2M\ln\left(\sigma^\ast_{\scriptscriptstyle
U}-\sigma_{\scriptscriptstyle U}\right)
\lab{t(UV)}\eeq
and
\beq x(U,V)\sim 2M\ln\left(\sigma^\ast_{\scriptscriptstyle
U}-\sigma_{\scriptscriptstyle U}\right)\;.
\lab{x(UV)}\eeq
Notice that $t\to +\infty$ and $x\to -\infty$ for $U\to
U^\ast$. The worldlines of the observers belonging to the
optical frame, $x=\mbox{const}$, tend to lie along the event
horizon, as one can also see from Eq.\ (\ref{dU/dV}), which
implies $\dd U/\dd V\to 0$ as $U\to U^\ast$. All these
features are very satisfactory from the point of view of a
smooth extension of the Schwarzschild rest frame inside the
star, and are in full agreement with the general results
obtained in Sec.\ \ref{4}.


\section{Hawking effect}

\lab{5}

\begin{figure}
\centerline{\epsfysize=7cm \epsfbox{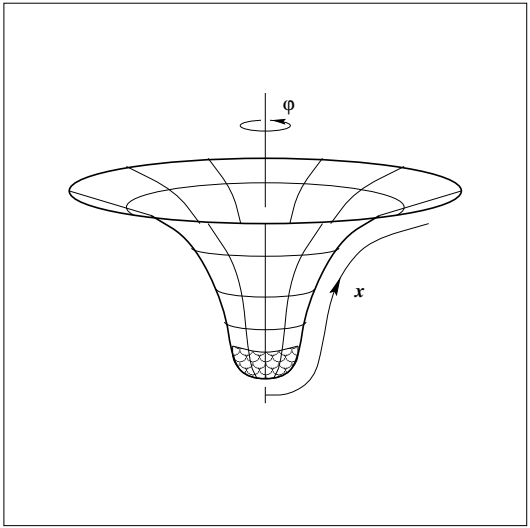}}
\vskip.4cm
\caption{Embedding diagram of the $\theta=\pi/2$ section
of optical space at an early stage of collapse.  Points in
the shadowed area are located inside the star.}
\label{fig8}
\end{figure}

\begin{figure}
\centerline{\epsfysize=7cm \epsfbox{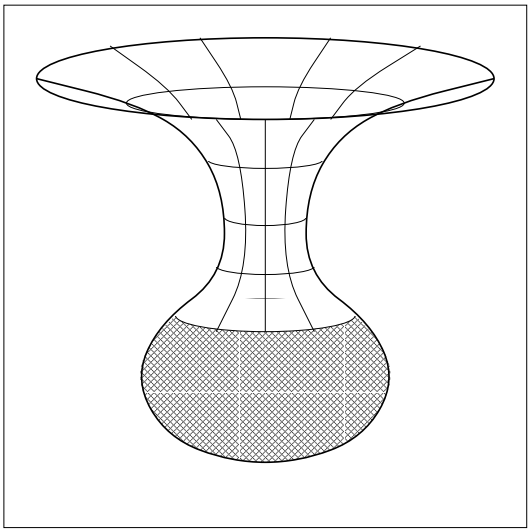}}
\vskip.4cm
\caption{The shape of the $\theta=\pi/2$ section of
optical space when $3M>R(t)>9M/4$.}
\label{fig9}
\end{figure}

\begin{figure}
\centerline{\epsfysize=7cm \epsfbox{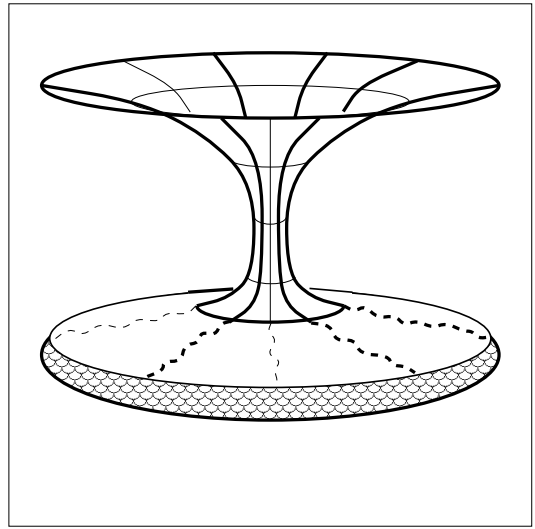}}
\vskip.4cm
\caption{When $R(t)<9M/4$, no faithful embedding is
possible in the three-dimensional Euclidean space. The star
escapes to $x\to -\infty$ at an increasing speed. The
optical radial distance between the centre $r=0$ and the
surface $r=R(t)$ tends to the finite value
$v_{\scriptscriptstyle P}-v_{\scriptscriptstyle H}$ when
$t\to +\infty$, while the area of the surface increases
without bound.}
\label{fig10}
\end{figure}

We now want to figure out how collapse looks like in the
optical geometry. For this purpose it is convenient to
consider the embedding diagrams of an equatorial plane at
different values of $t$ (see Ref.\ \cite{ksa} for a general
discussion of embedding diagrams). When $R>3M$, there are no
qualitative differences with respect to the conventional
description (see Fig.\ \ref{fig8}). However, as $R$ becomes
smaller than $3M$, a ``throat'' develops in the external
space, in correspondence with $r=3M$, while the surface of
the star expands progressively (Figs.\ \ref{fig9} and
\ref{fig10}), escaping to $x\to -\infty$ with the asymptotic
law (\ref{Main}). Thus, ``collapse'' actually corresponds,
in the optical geometry, to a sort of expansion into a space
that is created by the process itself. This picture is less
absurd than it may seem, if one remembers the operational
meaning of optical distance outlined in Sec.\ \ref{1}.
Consider an observer standing at a large value of $r$, who
sends light signals on a mirror that has been previously
placed at the centre of the star, and then defines his
distance from the centre simply in terms of the lapse of
Killing time $t$ taken by the round trip. Since the time
delay becomes progressively larger, he will deduce that a
collapsing star recedes from him at an increasing speed.
Because of spherical symmetry, the only geometrical picture
consistent with this description is the one of Figs.\
\ref{fig8}--\ref{fig10}, where the star expands into a
``lower space'' that is created in the course of collapse.

If a quantum field is present, this dynamical process
disturbs its modes analogously to what happens when there is
a moving boundary. In fact, in the two-dimensional section
shown in Fig.\ \ref{fig6}, the situation is {\em exactly\/}
the same as if there were a moving boundary, that
accelerates asymptotically towards the speed of light, with
the law (\ref{Main}). In Minkowski spacetime, this is known
to lead, at late times $t$, to a thermal flux of radiation
with temperature $(8\pi M)^{-1}$ \cite{boundary} (see also
Ref.\ \cite{bd}, pp.\ 102--109 and 229). Given the identity
between the $(t,x)$ part of the line element of
$\tilde{g}_{ab}$ with the one of a two-dimensional Minkowski
spacetime, one expects therefore that even in the late
stages of collapse there must be a flux of radiation at
temperature $T_{\scriptscriptstyle H}=(8\pi M)^{-1}$. This
is precisely the quantum emission found by Hawking
\cite{hawking,bd,boundary}, whose existence follows
therefore in a natural way from the picture of collapse
based on the optical geometry. It is remarkable that,
although the moving boundary analogy has often been used in
the literature \cite{boundary}, optical geometry gives a
very simple physical explanation of {\em why\/} it works:
Essentially, because gravitational collapse {\em is\/} a
particular case of a moving boundary.\footnote{Although the
optical space $\cal S$ has not a boundary in the technical
sense of the topology of manifolds, the centre of the star
works in the same way as far as fields are concerned,
because the boundary conditions at $r=0$ coincide with those
of perfect reflection.}

Actually, the analogy with a moving boundary in Minkowski
spacetime fails to hold exactly when the angular dimensions
are taken into account, because the field propagates on a
non-Euclidean geometry. However, this has the only effect of
distorting the spectrum of the emitted radiation, according
to the way waves are scattered by the curvature of space.
Such a correction is well-known in the theory of the Hawking
effect \cite{hawking,bd}.

In this picture the event horizon does not play any role in
deriving Hawking's radiation. In this respect, it is worth
reminding that in the optical geometry the horizon is very
much alike to a portion of the asymptotic null infinity. To
claim that it has anything to do with the thermal flux from
black holes would be exactly analogous to saying that the
future null infinity is relevant for the radiation produced
by a moving boundary in Minkowski spacetime. Thus,
explanations of the effect based on quantum tunneling
mechanisms appear rather implausible from the perspective of
optical geometry.

A similar situation occurs for black hole entropy. In the
optical geometry, the celebrated relation $S=A/4$ has no
direct meaning, because the horizon has an infinite area
according to the metric $\tilde{h}$. Indeed, the horizon is
located at $x\to -\infty$ in the optical space. If some
notion of entropy can be introduced, it can therefore be
associated only with the region $r>2M$. One possibility is
to attribute the entropy entirely to the Hawking radiation,
in the following way. Let us consider Fig.\ \ref{fig10}
again, where the space around a collapsing star at
sufficiently late times is visualized as made of a vast
``lower'' region connected to the usual ``upper'' one
through a mouth at $r=3M$. To observers living at large
values of $r$, the $r=3M$ surface looks like the boundary of
a three-dimensional cavity containing thermal black body
radiation at the temperature $T_{\scriptscriptstyle H}$.
(One may even consider the $r<3M$ region as analogous to a
Kirchhoff cavity in ordinary thermodynamics.) Whenever a
small amount of energy $\Delta E$ escapes to infinity as
Hawking's radiation, or is added to the collapsing star in
the form of accreted matter, the total entropy of the
radiation contained in the cavity is modified by the amount
$\Delta S=\Delta E/T_{\scriptscriptstyle H}$. Since $\Delta
E$ coincides with the change in the mass, as measured from
infinity, we recover the Beckenstein-Hawking expression
$S=4\pi M^2$ \cite{hawking,boundary,bekenstein} in its
differential form. Thus, optical geometry suggests that one
should regard the so-called black hole entropy as being
actually associated with the Hawking radiation surrounding
the collapsing star.\footnote{That there must be a trapping
effect on waves (usually attributed to the reflection off an
effective potential barrier) is rather evident from the
embedding diagram of Fig.\ \ref{fig10}. This phenomenon, and
its relevance for the study of long-lived gravitational-wave
modes, has been discussed in some detail in Ref.\
\cite{aabgs}. In the present context, it is responsible for
the persistence of a consistent amount of radiation in the
region with $r<3M$.} This interpretation contrasts with a
viewpoint often expressed, according to which $S$ is a
property of spacetime, and is similar to others that
attribute entropy to a ``thermal atmosphere'' of the black
hole \cite{york}.

Until now, we have assumed that the quantum field is a {\em
test\/} one, i.e., that it does not affect the background
spacetime. Clearly, this approximation becomes invalid at
late times, when the amount of energy carried away by the
Hawking radiation that leaks through the neck at $r=3M$
becomes a non-negligible fraction of the mass $M$ of the
star. Let us then sketch qualitatively how back-reaction
modifies the picture of collapse suggested by the optical
geometry. Roughly, the main effect of Hawking's emission is
to decrease the value of $M$. This has essentially three
consequences on the diagram of Fig.\ \ref{fig10}: First, it
decreases the size of the mouth at $r=3M$; second, it
increases the curvature of the optical space at the mouth
and in the lower region; third, it increases the value of
the Hawking temperature $T_{\scriptscriptstyle H}$. Thus, as
the process goes on and more radiation is able to escape to
$r\to +\infty$, the geometry in the region $r>3M$ of space
becomes closer and closer to the Euclidean one, while the
throat at $r=3M$ shrinks down, becoming progressively
sharper. The region $r<3M$ has larger and larger negative
curvature and contains, at large negative values of $x$,
matter that escapes to $x\to -\infty$ with increasing
acceleration.

Although this description is purely qualitative and is not
based on an explicitly constructed model, it gives us
important clues about the issue of the final state of a
collapsing star, when Hawking's radiation is taken into
account. First of all, it implies that the black hole, in
the strict sense of the region beyond the future event
horizon ${\cal H}^+$, does not form. This is evident in the
optical geometry point of view, where ${\cal H}^+$
corresponds to $x\to -\infty$ and $t\to +\infty$. In the
usual language, one would say that radiation makes the value
of $M$ decrease, which in turn makes the horizon shrink, and
eventually reduce to a point before the star could cross it.
Thus, the very existence of the Hawking effect would prevent
the formation of black holes by collapse.\footnote{This was
suggested already in the early years following Hawking's
discovery \cite{gerlach}, but the idea has never been
pursued further, at least to the authors' knowledge.}

If this is the case, what are then we left with in the limit
$t\to +\infty$? A straightforward extrapolation of the
process that we have just described suggests that, for $M\to
0$, the throat at $r=3M$ pinches off, leaving two spaces ---
one Euclidean, the other with infinite negative curvature
--- with just one point in common. This is clearly a
degenerate, and highly implausible, situation; one would
rather think that the process stops when the throat reaches
a critical size (perhaps at the Planckian scale), because of
as yet unknown physical processes. Such a situation
corresponds to the hypothesis of remnants in the common
description \cite{acn,giddings}.

Remnants have originally been introduced as a possible
resolution of the information paradox
\cite{giddings,infopar}. However, the viability of this
hypothesis has sometimes been questioned, because the small
scale and mass of remnants would allow them to store very
little information \cite{info}. Roughly, the argument is the
following. A physical system with size $l$, total energy
$E$, and Hamiltonian bounded from below has a number of
states of the order of $El$; thus, it can contain a maximum
information which is also of order $El$. The status of this
claim is rather controversial \cite{bounds}. However, even
assuming that there is indeed such an information bound, it
would hardly represent a difficulty in optical geometry,
where the ``remnant'' is actually an enormously vast region,
because ordinary distances are rescaled by a huge
factor.\footnote{This is similar to what happens in the
``cornucopion'' scenario of dilaton gravity coupled to
electromagnetism \cite{cornucopions} (see Ref.\
\cite{cornucrit} for criticisms of this model). However, the
internal geometries and the physical contexts are very
different in the two cases. The possibility that information
be stored inside a large region delimited by a small neck
has been discussed in general by Giddings \cite{giddings}.}
Even if $l$ should represent the size of the remnant as seen
from the exterior, as suggested by Bekenstein \cite{info},
the bound would still be circumvented thanks to the
existence, in the lower space, of an enormous supply of
negative energy associated with the field $\Phi$.


\section{Conclusions and outlooks}

\lab{6}

We have constructed the optical geometry for a spherically
symmetric collapsing body. The procedure adopted is a simple
extension of the technique used when defining optical
distance operationally in a static spacetime. Essentially,
for any spacetime event $P$, one considers the in- and
out-going spherical light-fronts that cross at $P$. These
correspond to well-defined values of advanced and retarded
time $v$ and $u$, that can be read at infinity as the affine
parameters along the null generators of ${\cal I}^-$ and
${\cal I}^+$. The event $P$ is then labeled by $v$ and $u$,
in addition to the angular coordinates, and can also be
identified by a timelike coordinate $t:=(v+u)/2$ and a
spacelike one, $x:=(v-u)/2$. The optical metric is then
defined as the only one which is conformal to $g_{ab}$ ---
the metric of spacetime in general relativity --- and which
reproduces the two-dimensional Minkowskian line element on
$\theta=\mbox{const}$, $\varphi=\mbox{const}$ sections.
Outside the collapsing star, $g_{ab}$ has the Schwarzschild
form and $x$ coincides with the Regge-Wheeler ``tortoise''
coordinate, thus giving the well-known optical metric of an
empty, spherically symmetric spacetime. However, our
construction allows one to extend the optical geometry
smoothly inside the star.

Associated with the optical space are the optical frame
$n^a$, made of the observers at $x=\mbox{const}$, and the
scalar potential $\Phi$, essentially the logarithm of the
conformal factor that links $g_{ab}$ and $\tilde{g}_{ab}$.
These two concepts are related to each other in the
following sense. The optical frame defines a notion of
``rest in the gravitational field,'' thus the
four-acceleration $n^b\nab_bn_a$ of the observers can be
identified with the gravitational acceleration, changed of
sign. Since $n^b\nab_bn_a$ is equal to the spatial gradient
of $\Phi$ in the optical frame, see Eq.\ (\ref{A}), $\Phi$
can be thought of as a covariant generalization of the
gravitational potential. This allows one to give a precise
meaning to the notion of gravitational field inside a
collapsing object. It must be noted that a gravitational
field so defined, although analogous to the same Newtonian
concept, nevertheless differs from it in some essential
details. As an example, consider a collapsing spherical
shell. It is obvious from Fig.\ \ref{fig7} that
$n^b\nab_bn_a\neq 0$ everywhere, thus leading to the
conclusion that the gravitational field is nonvanishing {\em
inside\/} the shell, as well as outside. This contrasts with
intuition shaped after Newton's theory, according to which
the gravitational field inside the shell should be zero.
However, this difference is not surprising if one considers
that $\Phi$ does not obey the Poisson equation, but a
nonlinear generalization of it, as can be easily checked on
replacing $g_{ab}=\ee^{2\Phi}\tilde{g}_{ab}$ in the trace of
the Einstein equation.

Although we have focused our treatment on the case of a star
collapsing in empty space, generalizations to, e.g., the
collapse of electrically charged bodies are straightforward
(see, e.g., Ref.\ \cite{lrs} for a configuration with
extremal charge, $Q^2=M^2$). In fact, the construction
presented in Sec.\ \ref{2} makes a heavy use of the null
structure on ${\cal I}^\mp$ in defining the coordinates $v$
and $u$, but seems rather general otherwise. It would be
important to understand whether a similar technique could be
used to define optical geometry in still more general
situations --- for example, for non-spherically symmetric
collapse.

We have seen that, in optical geometry, gravitational
collapse corresponds to a very fast expansion of the star
into a ``lower space,'' with the centre receding from the
optical observers $n^a$ at a speed that approaches
exponentially the speed of light. This process excites the
modes of a quantum field, as it happens in the presence of a
moving boundary, leading to the production of Hawking
radiation. Thus, optical geometry provides one with a
physical origin of the formal ``moving mirror analogy'' of
the Hawking effect.

The issue of the final state of black hole evaporation is
also clarified by the use of optical geometry. There are
essentially two possibilities, in both of which the black
hole does not form, strictly speaking. Either the
evaporation process continues until one remains with flat
space and a (physically unaccessible) infinitely warped
``lower space'' or, perhaps more likely, the process stops
at some scale leaving an enormously large remnant. In both
cases there is no information paradox. Of course, since the
issue involves distances of the order of the Planck length,
a definitive answer is beyond the limits of applicability of
present-day physics. Nevertheless, it would be useful to get
a more detailed insight into the evaporation process by
reformulating simple models that include back-reaction (see,
e.g., Ref.\ \cite{balbinot}) in the language of optical
geometry.

Of course, all our conclusions apply to black holes deriving
from collapse. However, it is not difficult to extend them
to eternal black holes, simply studying the quantum field
theory on the optical spacetime associated with the
Schwarzschild solution. In this case there is no moving
boundary, and the optical metric looks (as far as the $t$
and $x$ coordinates are concerned) exactly like the
Minkowski one. In particular, the observers at
$x=\mbox{const}$, belonging to the optical frame, correspond
to the Minkowskian inertial observers. Thus, one expects
that they should register no particles, provided that a
condition of ``no incoming radiation'' is imposed for $t\to
-\infty$.  This is precisely what happens in the so-called
Boulware (or Schwarzschild) state $|0_{\scriptscriptstyle
B}\rangle$ \cite{boulvac}. The optical geometry viewpoint
provides therefore support for regarding
$|0_{\scriptscriptstyle B}\rangle$, rather than the
Hartle-Hawking (or Kruskal) and the Unruh states
$|0_{\scriptscriptstyle H}\rangle$ and
$|0_{\scriptscriptstyle U}\rangle$ (see Ref.\ \cite{bd},
pp.\ 281--282), as describing the quantum vacuum around an
eternal black hole.

The Boulware state is often considered pathological because
the expectation value of several physical quantities
diverges at the horizon \cite{candelas}. For example, for
the stress-energy-momentum tensor operator of a massless
scalar field $\phi$ one has, after renormalization, that
$\langle 0_{\scriptscriptstyle
B}|T_{ab}|0_{\scriptscriptstyle B}\rangle_{\rm ren}$ depends
on $r$ as $(1-2M/r)^{-1}$ for $r\to 2M$. It is remarkable
that, in the optical spacetime $({\cal M},\tilde{g}_{ab})$,
such a factor is exactly canceled out, because $\phi$ is
conformally transformed in $\tilde{\phi}=\ee^\Phi\phi$, and
$\langle 0_{\scriptscriptstyle
B}|\widetilde{T}_{ab}|0_{\scriptscriptstyle B}\rangle_{\rm
ren}$ turns out to be finite everywhere, representing just
the vacuum polarization due to curvature. A similar
situation occurs when considering the response function
$\Pi(\omega|r)$ for an ideal static detector in the Boulware
state. In the spacetime $({\cal M},g_{ab})$ one has, near
the horizon \cite{candelas},
\beq \Pi(\omega|r)\sim -{1\over 1-2M/r}{\omega\over
2\pi}\,\Theta(-\omega)\;,
\eeq
where $\Theta$ is the step function. However, after the
conformal rescaling the response is given by
\beq \widetilde{\Pi}(\omega|r)\sim -{\omega\over
2\pi}\,\Theta(-\omega)\;,
\eeq
which not only is finite, but is also the answer one would
expect in a proper ``vacuum.'' Thus, the pathologies of the
Boulware state are removed by the conformal transformation
to the optical spacetime, in which $|0_{\scriptscriptstyle
B}\rangle$ behaves as a satisfactory quantum vacuum.


\acknowledgements

It is a pleasure to thank Stefano Liberati for many helpful
discussions. Part of this work was done while SS and JA were
at the Department of Astronomy and Astrophysics of Chalmers
University. SS was also partially supported by the
Interdisciplinary Laboratory of SISSA and by ICTP.


\section*{Appendix\ A:  Proof of Eq.\ (\ref{A})}

\def\theequation{A.\arabic{equation}}
\setcounter{equation}{0}

>From the definition (\ref{Phi}) of $\Phi$ it follows that
the vector field $n^a$ can be expressed as
$n^a=\ee^{-\Phi}\eta^a$.  Thus,
\beq n^b\nab_b n_a=\ee^{-2\Phi}\left(\eta^b\nab_b\eta_a-
\eta_a\eta^b\nab_b\Phi\right)\;.
\lab{a1}\eeq
Using now the relation
\beq \nab_a\eta_b+\nab_b\eta_a={1\over 2}\,g_{ab}\nab_c
\eta^c\;,
\lab{cKe}\eeq
valid for any conformal Killing vector field $\eta^a$ (see
e.g. Ref.\ \cite{wald}, pp.\ 443--444), we can write
\beq \eta^b\nab_b\eta_a=\ee^{2\Phi}\nab_a\Phi+{1\over 2}\,
\eta_a\nab_b\eta^b\;.
\lab{first}\eeq
Furthermore,
\beq \eta^b\nab_b\Phi={\eta^b\eta^c\nab_b\eta_c\over\eta_a
\eta^a}={\eta^b\eta^c\nab_{(b}\eta_{c)}\over\eta_a\eta^a}=
{1\over 4}\,\nab_b\eta^b\;.
\lab{second}\eeq
Substituting Eq.\ (\ref{second}) into Eq.\ (\ref{first}) and
then the latter into Eq.\ (\ref{a1}) we obtain, after some
trivial algebra, Eq.\ (\ref{A}).


\section*{Appendix B:  Physical meaning of the surface
gravity in optical geometry}

\def\theequation{B.\arabic{equation}}
\setcounter{equation}{0}

Here we show that, for a static black hole, the surface
gravity coincides with the magnitude of the gravitational
pull on the natural observers for $t\to +\infty$, as
measured with respect to the optical metric. A convenient
expression for the surface gravity $\kappa$ is
\beq \kappa=\lim\left[-{\left(\eta^a\nab_a\eta_c\right)
\left(\eta^b\nab_b\eta^c\right)\over \eta_d\eta^d}
\right]^{1/2}\;,
\lab{kappa}\eeq
where ``$\lim$'' stands for the limit as the horizon is
approached (see Ref.\ \cite{wald}, p.\ 332). From the
Killing equation and the definition of $\Phi$ it follows
that $\eta^a\nab_a\eta_c=
\nab_c\left(-\eta^a\eta_a\right)/2=
\ee^{2\Phi}\nab_c\Phi$.  Substituting into Eq.\
(\ref{kappa}) we obtain
\beq \kappa=\lim\left(\tilde{g}^{ab}\nab_a\Phi\nab_b
\Phi\right)^{1/2}=
\lim\left(\tilde{h}^{ab}\widetilde{D}_a\Phi\widetilde{D}_b
\Phi\right)^{1/2}\;,
\lab{B2}\eeq
where $\widetilde{D}$ is the Riemannian connection
associated with $\tilde{h}_{ab}$, and we have used the
stationarity property in the form $\eta^a\nab_a\Phi=0$. By
Eq.\ (\ref{A}), the last term in Eq.\ (\ref{B2}) is just the
value of the gravitational pull on the observers of the
optical frame.

This proof can be generalized without any difficulty to the
case of a stationary black hole, simply by replacing
$\eta^a$ with a Killing vector field which is normal to the
horizon.


\end{document}